\documentclass{article}
\usepackage{arxiv}

\usepackage{graphicx}
\usepackage[numbers]{natbib}
\usepackage{doi}

\title{Generating Photon Pairs in a hybrid Silicon-BTO Platform}

\author{ \href{https://orcid.org/0000-0000-0000-0000}{\includegraphics[scale=0.04]{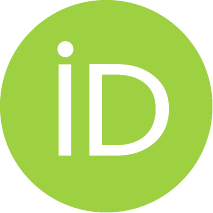}\hspace{1mm}D. Marchant}\\
    Quantum Engineering Centre for Doctoral Training \\
	Quantum Engineering Technology Labs \\
    University of Bristol, UK \\ 
	\texttt{d.marchant@bristol.ac.uk} \\
	\And
	\href{https://orcid.org/0000-0000-0000-0000}{\includegraphics[scale=0.04]{orcid.pdf}\hspace{1mm}I. Faruque} \\
	Quantum Engineering Technology Labs \\
    University of Bristol, UK \\ 
	\texttt{imad.faruque@bristol.ac.uk} \\
	\And
	\href{https://orcid.org/0000-0000-0000-0000}{\includegraphics[scale=0.04]{orcid.pdf}\hspace{1mm}J. Barreto} \\
	Quantum Engineering Technology Labs \\
    University of Bristol, UK \\ 
	\texttt{g.barreto@bristol.ac.uk} \\
}

\renewcommand{\headeright}{}
\renewcommand{\undertitle}{}
\renewcommand{\shorttitle}{Generating Photon Pairs in a hybrid Silicon-BTO Platform}

\hypersetup{
pdftitle={Generating Photon Pairs in a hybrid Silicon-BTO Platform},
pdfsubject={quant-ph},
pdfauthor={D. Marchant, I. Faruque, J. Barreto},
pdfkeywords={SFWM, BTO, pair generation, hybrid platform},
}

\begin{document}
\maketitle

\begin{abstract}
Here we show photon pair generation from ring resonator and waveguide structures in a hybrid silicon-BTO on insulator platform with a pulsed pump. Our analysis of single photon and coincidence generation rates show that Spontaneous Four-Wave Mixing is comparable to that expected from SOI devices of similar characteristics and find a $\gamma_{eff}$ of (14.7 $\pm$ 1.3) and (2.0 $\pm$ 0.3)  MHz/mW$^{2}$ for ring resonator and waveguide structures respectively.       
\end{abstract}


\section{Introduction}

Photon pair sources are a key building block for linear optical quantum computing \cite{Knill2001-ru, RevModPhys.79.135} and quantum communication \cite{qcomm, Ma2008-yf}. 
On-chip quantum information processing requires the capability of producing photon pairs with high brightness and high speed modulation. A hybrid Si-BTO platform was shown by Eltes et al. \cite{8613782} to fulfill the latter and we show here that it also fulfills the former requirement. This is in addition to cryogenic compatibility \cite{Eltes2020}, a requirement for full integration of integrated photonic circuits with SNSPDs, complementary metal-oxide-semiconductor (CMOS) compatibility \cite{8268454} and low loss \cite{Eltes2016}.

Integrated photon pair sources can be realised using spontaneous parametric down conversion (SPDC), a $\chi^{(2)}$ process or SFWM, a $\chi^{(3)}$ process \cite{Caspani2017}. In these two processes, one (in SPDC) or two (in SFWM) photons from the pump source are annihilated and two photons, a signal photon and idler photon, are generated at energy matched frequencies detuned from the pump. The frequency detuning of the generated photons allows them to be spectrally filtered from the pump photons. Although the $\chi^{(2)}$ coefficient of a material is typically an order of magnitude larger \cite{Boyd}, we have more freedom in exploiting the $\chi^{(3)}$ coefficient as it does not require a wide transparency window. In this letter we focus on the photon generation properties from SFWM. 

It is often desirable to be able to tune the effective refractive index of the waveguide for integrated photonics applications. For example, multiplexing schemes can be realised using Mach Zhender Interferemoters (MZIs) to route photons to different outputs by applying a phase shift to one arm. In materials that do not have a significant $\chi^{(2)}$ response, such as Si, this leaves either thermo-optic (TO) modulation \cite{COCORULLO199819}, plasma dispersion (PD) modulation \cite{1073206} or electro-optic (EO) modulation via the $\chi^{(3)}$ Kerr-effect \cite{Chakraborty:20}. The TO effect is limited to kHz speeds and it's magnitude is reduced at cryogenic temperatures \cite{10.1063/1.4738989}. PD modulators also have poor cryogenic compatibility due to carrier freeze out at low temperatures. While Kerr-effect EO modulators are cryogenically compatible, the tuning efficiency is poor leading to large device footprints. Another option is to integrate Si waveguides with a material that is capable of EO modulation via the $\chi^{(2)}$ Pockel's effect which has been shown to achieve high speed modulation at cryogenic temperatures using compact devices \cite{8613782}. 

In this letter we show that as part of a $\chi^{(3)}$ photon pair source, the presence of a BTO layer does not appear to hinder the photon pair generation rate when comparing with brightness values from similar devices in silicon-on-insulator (SOI). We use the effective non-linearity from SFWM, $\gamma _{eff}$ to quantify this.

\section{Methods}

\begin{figure*}
\centering
\includegraphics[width=0.8\linewidth]{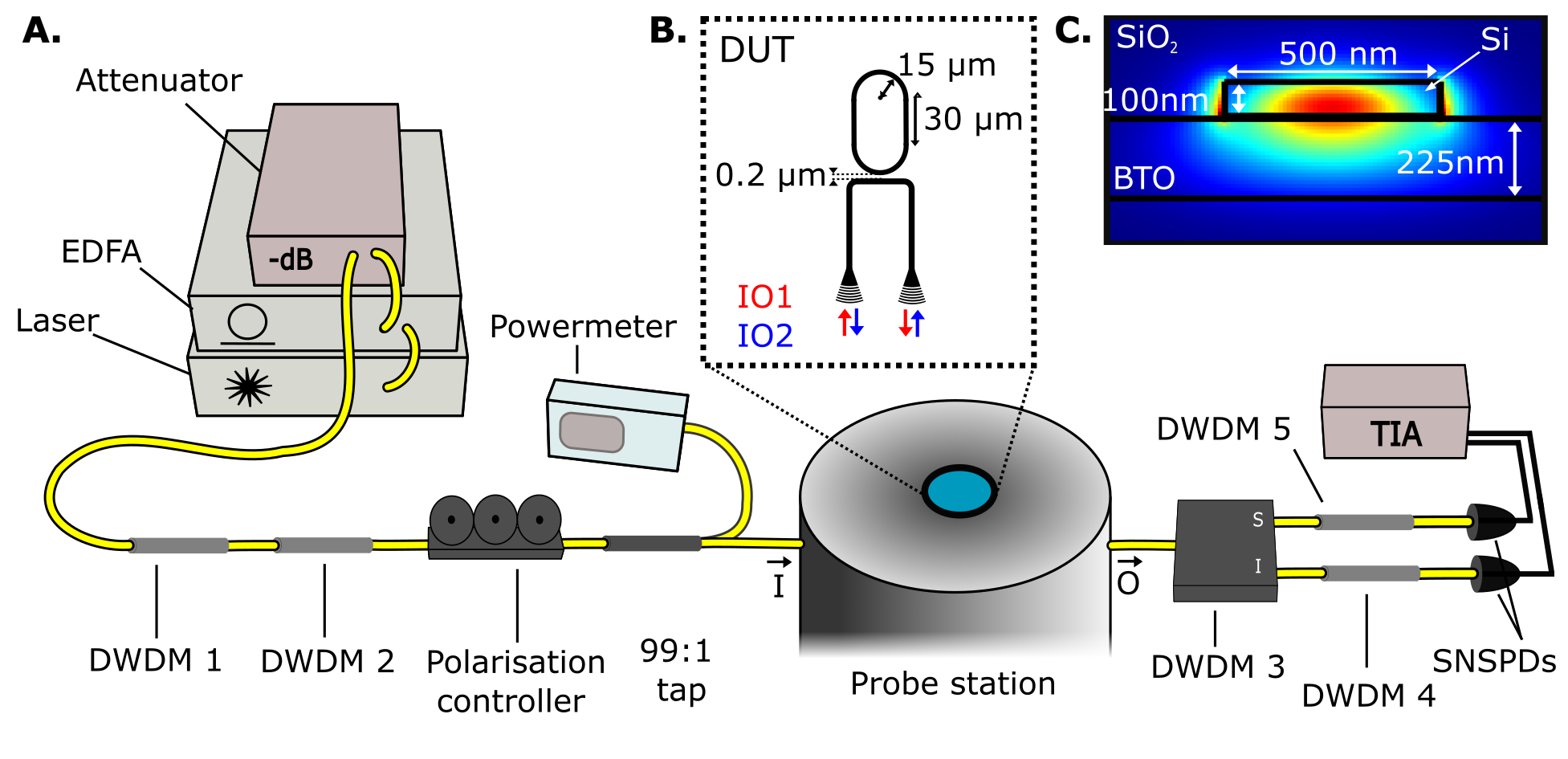}
\caption{\textbf{A.} Experimental setup used for characterising the device under test (DUT). \textbf{B.} Schematic of a DUT where the length of the bus waveguide section is 2.5 mm. The two possible grating coupler input and output configurations labeled IO1 \& IO2 \textbf{C.} Electric field distribution of the fundamental TE mode in a waveguide cross-section of the DUT. Details of all equipment are tabulated  in Table S3. }
\label{fig:schematic}
\end{figure*}

We tested the platform by characterizing a series of ring resonator devices with a 2.5 mm bus waveguide section. To compare with brightness values of waveguide and ring resonator devices in literature, we measured the device both on and off-resonance. This was done by controlling the temperature to isolate the contribution from the bus waveguide section and filter the signal and idler wavelengths each with a bandwidth of 0.5 nm and separated 4.8 nm from the pump filter.

The setup depicted in Figure \ref{fig:schematic} was employed to characterize the photon generation process in this hybrid platform. A pulsed laser centred at 1550.97 nm with a bandwidth of 0.52 nm and pulse width of 1 ps was amplified with an erbium-doped fibre amplifier (EDFA) to generate the strong pump for this process. This was then passed through a variable optical attenuator (VOA) to allow full tunability of the power down to -60 dB without changing the pulse properties. Two dense wavelength-division multiplexer (DWDM) filters with $\geq$80 dB of combined rejection band suppression were used to eliminate pump noise from the noise floor of the laser entering the signal and idler collection bandwidths. These were placed before a polarization controller to allow optimization of the inserted light for TE polarization that the grating couplers were designed for. A 99:1 beam splitter allowed monitoring of the power coupled to the device under test (DUT) through a V-groove optical fibre array mounted to one arm of the probe station. After the output from the probe station, a multi-channel DWDM module was used to de-multiplex the signal and idler frequencies and reject the pump. An extra single-channel DWDM was then placed on the signal and idler channels for additional filtering. This filtering after the chip also offered $\geq$80 dB of rejection band suppression for each channel. Finally, the two channels were routed through a fibre network to two PhotonSpot superconducting nano-wire single photon detectors (SNSPDs) connected to a time interval analyser (TIA).

\begin{figure*}
\centering
\includegraphics[width=1.0\textwidth]{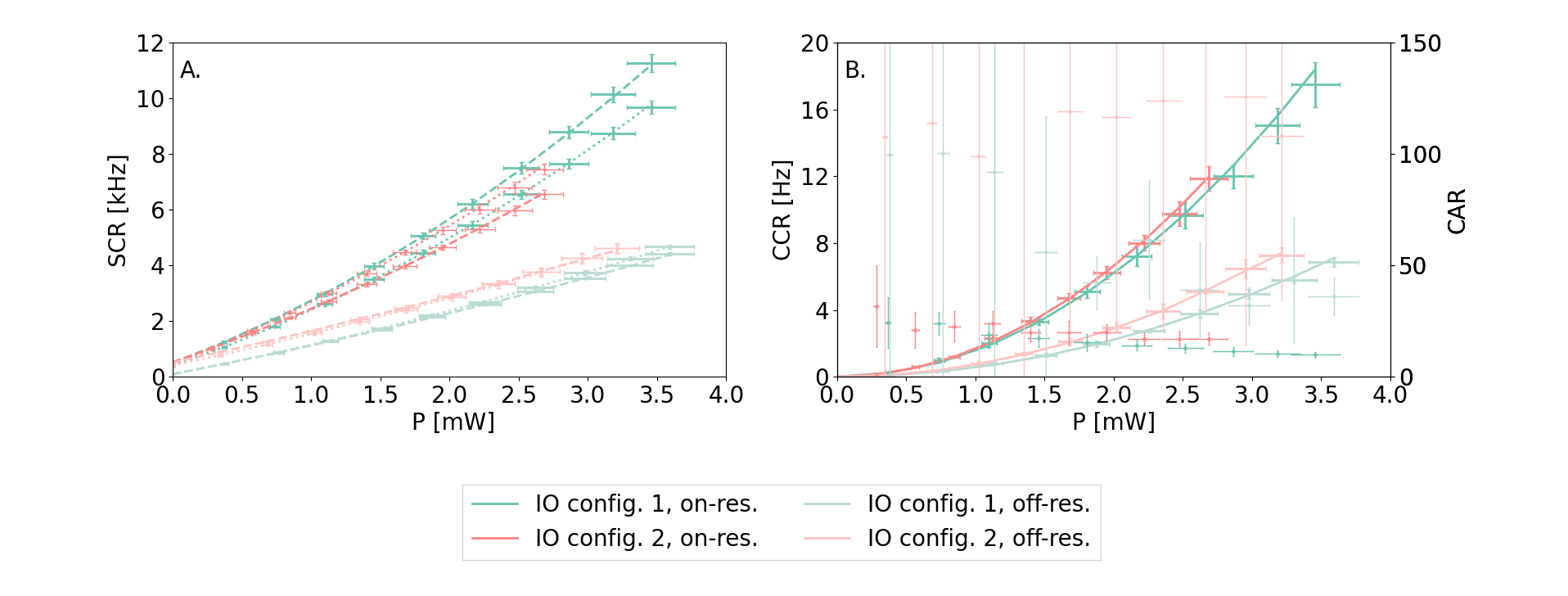}
\caption{\textbf{A.} shows the raw single count rates (SCR) as a function of pump power. The dashed and dotted lines are the fits of the signal and idler counts respectively. \textbf{B.}  shows the coincidence count rates (CCR) and CAR as a function of pump power. On- and off- resonance for the IO1 configuration are represented by darker and lighter green respectively. On- and off- resonance for the IO2 configuration are represented by darker and lighter red respectively. Raw count rates are represented by bold error bars and CAR values are represented by light error bars.}
\label{fig:main}
\end{figure*}

The coupling was optimised in both position and polarisation before each measurement with the pulsed laser and the total coupling loss from the pair of input and output grating couplers, $\eta _{coupling}$ measured. The chip was mounted to the sample stage of a Lakshore CPX probe station using thermal resin and the temperature of this stage was controlled using a resistive heater controlled by a temperature controller. The VOA was set to sweep over a set of attenuation values to yield the characteristic plots shown in Figure \ref{fig:main}. It was also necessary to limit the total time taken to carry out the measurements to limit the effect of coupling drift. The integration time was varied between 30 and 300 seconds depending on the level of attenuation. This was to ensure that the number of coincidence and single counts remained roughly constant and minimise the statistical variation across the measurements. 

To find $\gamma _{eff}$ for SFWM, a set of three quadratic equations can be used to model and fit a set of data which is taken by measuring the signal count rate, idler count rate and the signal-idler coincidence count rate as a function of power \cite{silicon_photonics_book}. For our devices, we found that a non-negligible imbalance of the input and output grating coupler efficiencies leads to an over or under estimation of $\gamma _{eff}$ when calculated in this way. This depends on which grating coupler is chosen to be the input or output and increases the uncertainty in $\gamma _{eff}$. We propose a method outlined the supplementary material that was inspired by the work of Sinclair et al. \cite{PhysRevApplied.11.044084}. This reduces the overall  uncertainty and was used to calculate the $\gamma _{eff}$ for our devices.  

The integration of counts for each attenuation value set on the VOA was repeated 10 times for each configuration and the standard error calculated from these repeated measurements were passed as a parameter in a least squares fit. Errors in the quadratic coefficients were then obtained from the covariance matrix of these fits.

\section{Results and Discussion}

Tables 1 and 2 show that the value we report for $\gamma_{eff}$ compares favourably with other waveguide and microring resonator sources where a similar method of excitation is used. We note that a large linear component to the signal and idler count rates is certainly a source of uncertainty in our values. Despite the filtering used having $\geq$80 dB of rejection band suppression on each channel, this still resulted in a small (of the order of $10^{2}$ Hz) amount of pump leakage. Suppressing this further would require longer integration times to detect a sufficient number of coincidence events which is already low due to high losses inherent in the system. $\eta _{coupling}$ was measured to be approximately 20 dB. The high insertion loss has been shown not to be due to the BTO layer\cite{Eltes2020}. It is likely due to a combination of the use of non-optimized grating couplers and a lack of pitch control on the optical arm of the probe station. The loss of the optical connection from where the probe station is located to the detector system was measured to be approximately 4 dB for each channel. Longer integration times also risk coupling drift over the series of measurements which arise due to thermal fluctuations in the environment.-

\begin{table*}[ht]
\centering
\begin{tabular}{ccccccc}
\hline
\textbf{Ref} & \textbf{Geometry [nm]}  & \textbf{Excitation} & \textbf{CAR} &  \textbf{L [mm]}  & \textbf{$\gamma _{eff}$ [MHz/mW$^{2}$]} \\
\hline
\citenum{PGR_USRN} & SiN[550x300] & CW & 7 & 10 & 0.009 \\
\citenum{PhysRevApplied.12.054029} & Si[500x220] & Pulsed & 14 & - & 1.1\\ 
\citenum{Kultavewuti:16} & AlGaAs[700x600] & Pulsed & 177 & 4.5 & 1.91 \\
This work & Si[100x220]/BTO[170] & pulsed & 108 & 2.5 & 2.0 $\pm$ 0.3\\ 
\citenum{Sharping:06} & Si[100x220] & Pulsed & 25 & 9.11 & 2.5 \\ 
\hline
\end{tabular}
\label{tab:wg_compare}
\caption{Comparison of waveguide photon pair source brightness measurements in order of magnitude of $\gamma _{eff.}$ reported.  The geometry is defined as material[waveguide width$\times$waveguide height] for strip waveguide geometry with /material[slab thickness] added for rib waveguide geometry. The excitation methods used were a continuous wave (CW) or pulsed laser pump.
}
\end{table*}

\begin{table*}[ht]
\centering
\begin{tabular}{ccccccc}
\hline
\textbf{Ref} & \textbf{Geometry}\textbf{[nm]} & \textbf{Excitation} & \textbf{CAR} & \textbf{C} \textbf{[$\mu$m]} & \textbf{Q [$\times$ 10 $^4$]}   &  \textbf{$\gamma _{eff}$} \textbf{[MHz/mW$^{2}$]} \\
\hline
\citenum{Wu:21} & SiN[8000x950] & CW &  1864 $\pm$ 571  & 386.4 & 100 & 1.03 $\pm$ 0.09  \\
\citenum{Clemmen:09} & Si[500x220] & CW & 30 & 42.7 & 1 & 1.9 \\ 
\citenum{Engin:13} & Si[500x220] & Pulsed; RB  & 602 $\pm$ 37 & 62.8 & 4 & 5.3 \\ 
\citenum{Fujiwara:17} & Si[400xs220] & CW & $>$350 & 62.8 & 10 & 5.8 \\ 
\citenum{Faruque:18} & Si[500x220] & Pulsed & $>$ 100 & - & 5 & 6.1 \\ 
\citenum{Ma:18} & Si[650x220]/Si[70] & Pulsed & 3,000 $\pm$ 500  & 62.8 & 9 & 14.6 \\
This work & Si[500x100]/BTO[225]& Pulsed & 32 $\pm$ 19  & 248.2 & 7 & 14.7$\pm$ 1.3 \\
\citenum{Zheng_2021} & Si[500x220] & Pulsed & - & 111.8 & 6  & 19.9 \\ 
\citenum{Silverstone2015QubitEB} & Si[500x220] & Pulsed & 10 & 94.2 & - & 30.6 \\ 
\citenum{Ma:17} & Si[650x220]/Si[70] & CW; RB & 12,105 $\pm$ 1,821  & 62.8 & 9 & 149 $\pm$ 6\\
\hline
\end{tabular}
\label{tab:mrr_compare}
\caption{Comparison of micro ring resonator photon pair source brightness measurements in order of magnitude of $\gamma _{eff.}$ reported. *In cases where a $\gamma _{eff}$ value was not directly reported, this was extrapolated using the maximum  coincidence count rate and corresponding value for power. The geometry is defined as material[waveguide width$\times$waveguide height] for strip waveguide geometry with /material[slab thickness] added for rib waveguide geometry. The excitation methods used were a continuous wave (CW) or pulsed laser pump with one scheme using additional reverse biasing (RB). Q is defined as the loaded quality factor. 
}
\end{table*}

Our reported value for coincidence-to-accidental ratio (CAR, details outlining our calculation of this is provided in the supplementary material) is relatively low compared to other values reported in Table 2. The main factor contributing to this is the high channel loss as this increases the probability of broken pairs contributing to the accidental counts. The large error in the CAR is due to the very small number of accidental counts observed at the lowest pump power. This was particularly apparent in the off-resonance case. A higher value with lower uncertainty could be achieved if coupling could be sustained at lower pump powers for a longer integration time of the order of several hours. The lowest pump power setting used was 0.3 mW with an integration time of 300 seconds. Ma et al. \cite{Ma:18} achieve a CAR of 2,873 using a comparable amount of pulsed pump power in the waveguide but a longer integration time of 3,000 seconds. They also report an approximately 7 dB better collection efficiency of photons generated in the waveguide which is likely a factor in this higher value.

We have demonstrated the photon generation via SFWM in these devices at room temperature. Given the results of recent studies in Si\cite{Feng:23} and the cryogenic compatibility of the electro-optic effect in these devices \cite{Eltes2020}, we expect good photon generation performance at cryogenic temperatures as well.

We have shown here that the effective non-linearity of a hybrid Si-BTO integrated photon pair source is comparable to an all-Si one and appears to provide an enhancement compared to some reported values \cite{PhysRevApplied.12.054029, Clemmen:09, Engin:13, Fujiwara:17, Faruque:18, Ma:18}. Together with the work of Eltes et al. \cite{8613782}, our results support the case for implementing this platform in scenarios where high speed modulation and high brightness are required.

\section{Conclusions}

In this letter we demonstrate that a hybrid Si-BTO platform shows similar or better performance in terms of SFWM brightness, to a SOI platform. We characterize the brightness of our source by analyzing coincidence and single count rates as a function of injected pump power which leads to a calculated $\gamma_{eff}$ of (14.7 $\pm$ 1.3) and (2.0 $\pm$ 0.3) MHz/mW$^{2}$ for the on- and off-resonance conditions respectively. Given that these devices were not optimized for photon generation, higher brightness could likely be achieved by fine-tuning the dispersion engineering of the waveguide geometry. The maximum CAR value of between 32$\pm$19 and 24$\pm$12 is reasonable compared to other values quoted in literature. However, a higher value with lower uncertainty could be achieved with more efficient coupling structures and lower loss detection channels. Future studies repeating this experiment at low temperatures could demonstrate the predicted cryogenic compatibility of these devices and the effect on source brightness. In summary, we conclude that the presence of a BTO layer in a hybrid Si-BTO device provides no drawbacks in terms of brightness of the photon pair source while providing EOM capabilities.

\section*{Acknowledgments}
D.M. would like to thank Ben Burridge for useful discussions related to this work and acknowledge the support of the QECDT through the EPSRC training grant, EP/LO15730/1. J.B. would like to thank Graham D. Marshall and Mark G. Thompson for useful discussions.  All the authors would like to thank Stefan Abel and Felix Eltes (Lumiphase AG) for supporting this work.

\appendix

\section{The Devices}

\begin{figure*}
\renewcommand{\thefigure}{S\arabic{figure}} 
\centering
\includegraphics[width=0.9\linewidth]{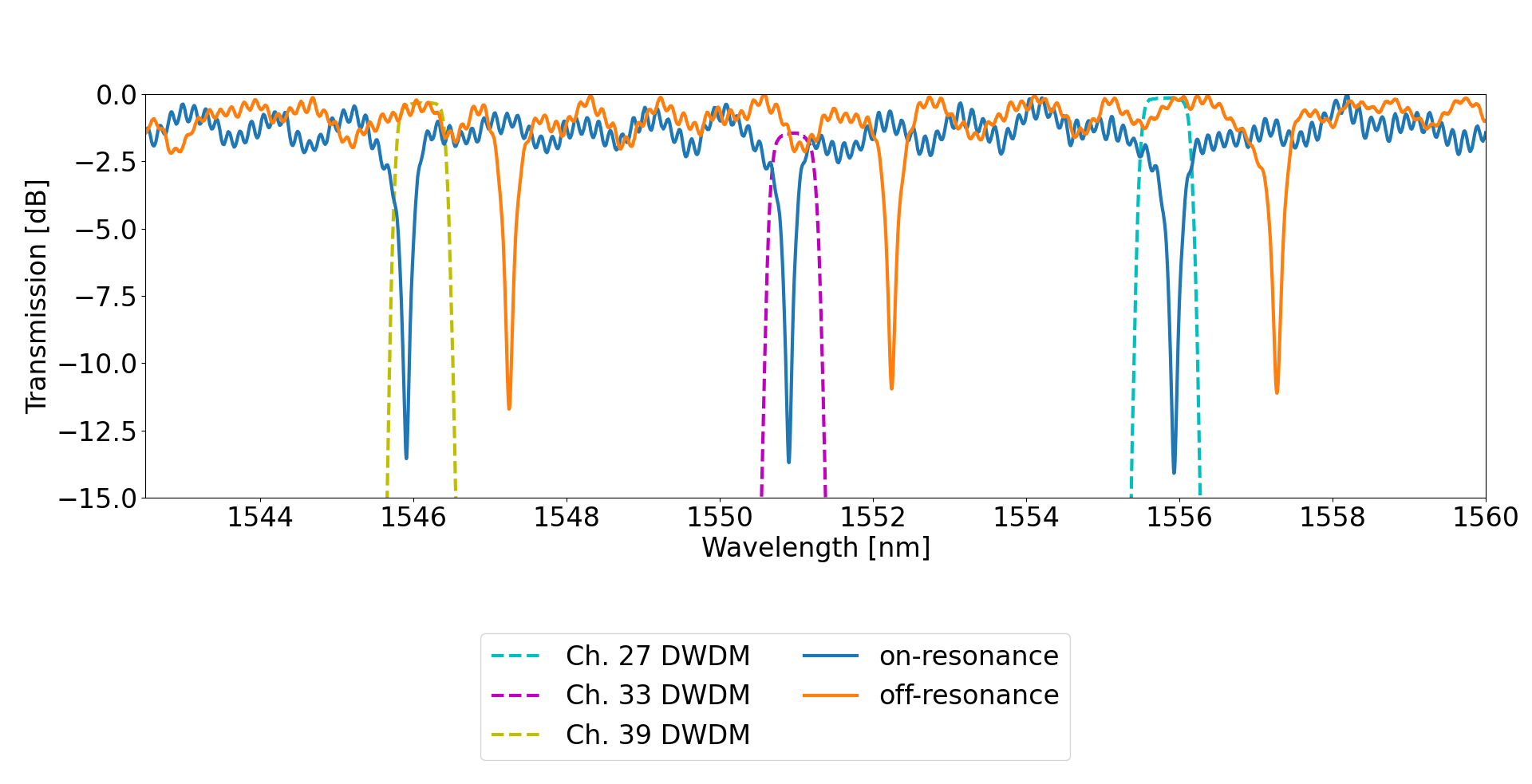}
\caption{Relative transmission of the microring resonator and DWDM filters that were used to determine $\gamma _{eff.}$.}
\label{fig:p_specs}
\end{figure*}

In Figure \ref{fig:p_specs} we show that when the microring resonator device is shifted to the on-resonance condition by adjusting the temperature of the sample stage to 295 K there are three consecutive resonances in good alignment with the three DWDM ITU channels that were used.  We also show that by adjusting the temperature to 325 K, the resonances are shifted completely outside of the channel bandwidths allowing us to ignore any contribution from the resonator.

Figure \ref{fig:gfit} was obtained by pumping the device on-resonance at the maximum pump power for a period of several minutes.  Plotting the histogram data and applying a Gaussian fit yields a FWHM of 1.2 ns.  In the case of perfect detectors (i.e. zero jitter) we would expect this to correspond to the coherence time of the generated photons.

\section{\label{metrics}Metrics}

The two key metrics we used to compare our devices are the source brightness and the coincidence-to-accidental ratio (CAR). 

In the case of SFWM, where two pump photons are involved in the process. brightness is defined as the rate of photon pairs produced per square unit of pump power.

\begin{figure}[ht]
\renewcommand{\thefigure}{S\arabic{figure}} 
\centering
\includegraphics[width=0.6\linewidth]{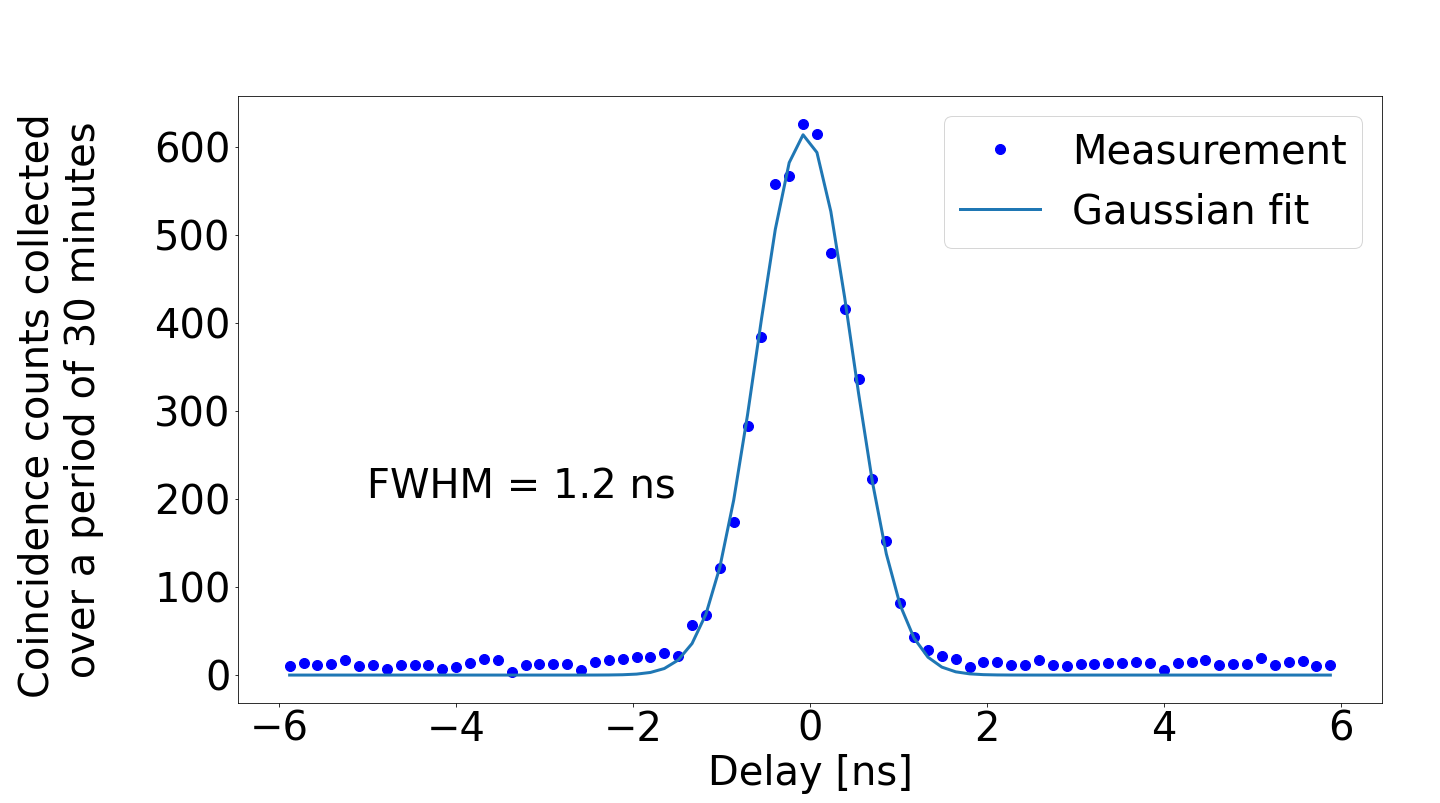}
\caption{Gaussian fit of coincidences collected from pumping the ring resonator in the on-resonance configuration shown in Figure \ref{fig:p_specs}.}
\label{fig:gfit}
\end{figure}

The CAR is the ratio of coincidence counts from SFWM events to accidental coincidence counts caused by broken pairs, pump noise and other detection events not caused by the experiment. For a pulsed laser, the histogram plot will consist a central peak which contains both coincidence and accidental events and adjacent peaks which contain only accidental events. The CAR is then calculated by subtracting the adjacent peak from the central peak and taking the ratio. We chose an integration window of 2 ns, which was guided by the fit in Figure \ref{fig:gfit}.

\begin{table*}[ht]
\renewcommand{\thetable}{S\arabic{table}} 
\centering
\begin{tabular}{cccc}
\hline
\textbf{ID} & \textbf{Description} & \textbf{Manufacture} & \textbf{Model Number}  \\
\hline
CPX & Cryogenic probe station & LakeShore & CPX-VF \\
TC & Temperature controller & LakeShore & 336 \\
TP & 50 MHz tunable pulsed laser &  PriTel & FFL-50MHZ \\
EDFA & High peak power optical fibre amplifier &  PriTel &  HPP-PMFA-21-10 \\
VOA & Variable optical attenuator &  HP &  8156A \\
PM & Optical powermeter & ThorLabs & PM100D \\
DWDM1 & ITU ch. 33 DWDM & Opneti & DWDM-1-1-C33-900-1-1-FA \\
DWDM2 & ITU ch. 33 DWDM & Opneti & DWDM-1-1-C33-900-1-1-FA \\
DWDM3 & ITU multi-ch. DWDM & Opneti & DWDM-D-2-16-D-21-900-1-1-FA \\
DWDM4 & ITU ch. 27 DWDM & Opneti & DWDM-1-1-C27-900-1-1-FA \\
DWDM5 & ITU ch. 39 DWDM & Opneti & DWDM-1-1-C39-900-1-1-FA \\
SNSPD & Closed-cycle cryostat SNSPD system & PhotonSpot & - \\
TIA & Time interval analyser & - & - \\
\hline
\end{tabular}
\caption{\label{tab:equipment} List of equipment shown Figure \ref{fig:schematic} }
\end{table*}

\section{\label{nonlinearity calculation}Effective non-linearity calculation}

In this section we provide a detailed outline of the method used to calculate the effective non-linearity of our device without knowing precisely the unbalanced losses of our grating couplers. First, we take the equations for the signal single count rate, $C_{s}$ the idler single count rate, $C_{i}$ and the coincidence count rate, $CC$ and rewrite them in terms of the collection efficiencies, $\eta_{1}$ and $\eta_{2}$ the input grating coupler efficiencies, $\eta_{GC_{A}}$ and $\eta_{GC_{B}}$  and the power set on the laser, $P_{laser}$

\begin{equation} 
\renewcommand{\thefigure}{S\arabic{figure}} 
\label{eq:4}
C_{s _{A,B}} = \eta_{1} \gamma _{eff} (\eta_{GC _{A,B}} P_{laser})^{2} + \eta_1\eta_{GC _{A,B}} P_{laser} + DC_{s}
\end{equation}

\begin{equation} 
\renewcommand{\thefigure}{S\arabic{figure}} 
\label{eq:5}
C_{i _{A,B}} = \eta_{2} \gamma _{eff} (\eta_{GC _{A,B}} P_{laser})^{2} + \eta_2\eta_{GC _{A,B}} P_{laser} + DC_{i}
\end{equation}

\begin{equation} 
\renewcommand{\thefigure}{S\arabic{figure}} 
\label{eq:6}
CC_{A,B} =  \eta_{1}\eta_{2} \gamma _{eff} (\eta_{GC _{A,B}} P_{laser})^{2} +ACC.
\end{equation}

where A, B denote the two possible input and output configurations of the grating couplers. By multiplying the coefficients of the quadratic terms in equation \ref{eq:4}, $a_{s}$ and equation \ref{eq:5}, $a_{i}$ then dividing the the quadratic term in equation \ref{eq:6}, $a_{si}$ we obtain

\begin{equation}
\renewcommand{\thefigure}{S\arabic{figure}} 
\label{eq:7} 
\frac{a_{s_{A,B}}a_{i_{A,B}}}{a_{si_{A,B}}} = \gamma _{eff} \eta_{GC _{A,B}} ^{2}
\end{equation}

Multiplying this together for both configurations yields

\begin{equation}
\renewcommand{\thefigure}{S\arabic{figure}} 
\label{eq:8} 
\frac{a_{s_{A}}a_{i_{A}}}{a_{si_{A}}} . \frac{a_{s_{B}}a_{i_{B}}}{a_{si_{B}}} = \gamma _{eff} ^{2} \eta_{GC _{A}}  ^{2} \eta_{GC _{B}}  ^{2} 
\end{equation}

Finally, given we know that the product of $\eta_{GC _{A}}$ and $\eta_{GC _{B}}$ is just the total coupling efficiency, $\eta_{coupling}$ for a grating coupler pair, we find that the effective non-linearity can be written in the following terms 

\begin{equation}
\renewcommand{\thefigure}{S\arabic{figure}} 
\label{eq:9} 
\gamma _{eff} = \sqrt{ \frac{a_{s_{A}}a_{s_{B}}a_{i_{A}}a_{i_{B}}}{a_{si_{A}}a_{si_{B}}\eta_{coupling}^{2}}}
\end{equation}

all of which can be determined from the fitting of equations \ref{eq:4}-\ref{eq:6} or measured directly in the case of $\eta_{coupling}$.

\bibliographystyle{unsrt}
\bibliography{references}  

\providecommand{\noopsort}[1]{}\providecommand{\singleletter}[1]{#1}%
\begin{thebibliography}{10}

\bibitem{Knill2001-ru}
E~Knill, R~Laflamme, and G~J Milburn.
\newblock A scheme for efficient quantum computation with linear optics.
\newblock {\em Nature}, 409(6816):46--52, January 2001.

\bibitem{RevModPhys.79.135}
Pieter Kok, W.~J. Munro, Kae Nemoto, T.~C. Ralph, Jonathan~P. Dowling, and G.~J. Milburn.
\newblock Linear optical quantum computing with photonic qubits.
\newblock {\em Rev. Mod. Phys.}, 79:135--174, Jan 2007.

\bibitem{qcomm}
Nicolas Gisin and R.~Thew.
\newblock Quantum communication.
\newblock {\em Nat. Photon.}, 1, 04 2007.

\bibitem{Ma2008-yf}
Xiongfeng Ma and Hoi-Kwong Lo.
\newblock Quantum key distribution with triggering parametric down-conversion sources.
\newblock {\em New J. Phys.}, 10(7):073018, July 2008.

\bibitem{8613782}
Felix Eltes, Christian Mai, Daniele Caimi, Marcel Kroh, Youri Popoff, Georg Winzer, Despoina Petousi, Stefan Lischke, J.~Elliott Ortmann, Lukas Czornomaz, Lars Zimmermann, Jean Fompeyrine, and Stefan Abel.
\newblock A batio3-based electro-optic pockels modulator monolithically integrated on an advanced silicon photonics platform.
\newblock {\em Journal of Lightwave Technology}, 37(5):1456--1462, 2019.

\bibitem{Eltes2020}
Felix Eltes, Gerardo~E. Villarreal-Garcia, Daniele Caimi, Heinz Siegwart, Antonio~A. Gentile, Andy Hart, Pascal Stark, Graham~D. Marshall, Mark~G. Thompson, Jorge Barreto, Jean Fompeyrine, and Stefan Abel.
\newblock An integrated optical modulator operating at cryogenic temperatures.
\newblock {\em Nature Materials}, 19(11):1164--1168, Nov 2020.

\bibitem{8268454}
F.~Eltes, M.~Kroh, D.~Caimi, C.~Mai, Y.~Popoff, G.~Winzer, D.~Petousi, S.~Lischke, J.~E. Ortmann, L.~Czornomaz, L.~Zimmermann, J.~Fompeyrine, and S.~Abel.
\newblock A novel 25 gbps electro-optic pockels modulator integrated on an advanced si photonic platform.
\newblock In {\em 2017 IEEE International Electron Devices Meeting (IEDM)}, pages 24.5.1--24.5.4, 2017.

\bibitem{Eltes2016}
Felix Eltes, Daniele Caimi, Florian Fallegger, Marilyne Sousa, Eamon O'Connor, Marta~D. Rossell, Bert Offrein, Jean Fompeyrine, and Stefan Abel.
\newblock Low-loss batio3--si waveguides for nonlinear integrated photonics.
\newblock {\em ACS Photonics}, 3(9):1698--1703, Sep 2016.

\bibitem{Caspani2017}
Lucia Caspani, Chunle Xiong, Benjamin~J Eggleton, Daniele Bajoni, Marco Liscidini, Matteo Galli, Roberto Morandotti, and David~J Moss.
\newblock Integrated sources of photon quantum states based on nonlinear optics.
\newblock {\em Light Sci Appl}, 6(11):e17100, June 2017.

\bibitem{Boyd}
Robert Boyd.
\newblock {\em Chapter 1. The Nonlinear Optical Susceptibility}, pages 1--65.
\newblock Academic Press, 12 2003.

\bibitem{COCORULLO199819}
Giuseppe Cocorullo, Francesco~G. {Della Corte}, Ivo Rendina, and Pasqualina~M. Sarro.
\newblock Thermo-optic effect exploitation in silicon microstructures.
\newblock {\em Sensors and Actuators A: Physical}, 71(1):19--26, 1998.

\bibitem{1073206}
R.~Soref and B.~Bennett.
\newblock Electrooptical effects in silicon.
\newblock {\em IEEE Journal of Quantum Electronics}, 23(1):123--129, 1987.

\bibitem{Chakraborty:20}
Uttara Chakraborty, Jacques Carolan, Genevieve Clark, Darius Bunandar, Gerald Gilbert, Jelena Notaros, Michael~R. Watts, and Dirk~R. Englund.
\newblock Cryogenic operation of silicon photonic modulators based on the dc kerr effect.
\newblock {\em Optica}, 7(10):1385--1390, Oct 2020.

\bibitem{10.1063/1.4738989}
J.~Komma, C.~Schwarz, G.~Hofmann, D.~Heinert, and R.~Nawrodt.
\newblock {Thermo-optic coefficient of silicon at 1550 nm and cryogenic temperatures}.
\newblock {\em Applied Physics Letters}, 101(4):041905, 07 2012.

\bibitem{silicon_photonics_book}
D.~Bonneau, J.~W. Silverstone, and M.~G. Thompson.
\newblock {\em Silicon Photonics III: Systems and Applications}.
\newblock Springer, 2016.

\bibitem{PhysRevApplied.11.044084}
Gary~F. Sinclair, Nicola~A. Tyler, D\"ond\"u Sahin, Jorge Barreto, and Mark~G. Thompson.
\newblock Temperature dependence of the kerr nonlinearity and two-photon absorption in a silicon waveguide at 1.55 $\ensuremath{\mu}\mathrm{m}$.
\newblock {\em Phys. Rev. Appl.}, 11:044084, Apr 2019.

\bibitem{PGR_USRN}
Ju~Won Choi, Byoung-Uk Sohn, George Chen, Doris Ng, and Dawn Tan.
\newblock Correlated photon pair generation in ultra-silicon-rich nitride waveguide.
\newblock {\em Optics Communications}, 463:125351, 05 2020.

\bibitem{PhysRevApplied.12.054029}
Imad~I. Faruque, Gary~F. Sinclair, Damien Bonneau, Takafumi Ono, Christine Silberhorn, Mark~G. Thompson, and John~G. Rarity.
\newblock Estimating the indistinguishability of heralded single photons using second-order correlation.
\newblock {\em Phys. Rev. Appl.}, 12:054029, Nov 2019.

\bibitem{Kultavewuti:16}
Pisek Kultavewuti, Eric~Y. Zhu, Li~Qian, Vincenzo Pusino, Marc Sorel, and J.~Stewart Aitchison.
\newblock Correlated photon pair generation in algaas nanowaveguides via spontaneous four-wave mixing.
\newblock {\em Opt. Express}, 24(4):3365--3376, Feb 2016.

\bibitem{Sharping:06}
Jay~E. Sharping, Kim~Fook Lee, Mark~A. Foster, Amy~C. Turner, Bradley~S. Schmidt, Michal Lipson, Alexander~L. Gaeta, and Prem Kumar.
\newblock Generation of correlated photons in nanoscale silicon waveguides.
\newblock {\em Opt. Express}, 14(25):12388--12393, Dec 2006.

\bibitem{Wu:21}
Kaiyi Wu, Qianni Zhang, and Andrew~W. Poon.
\newblock Integrated si3n4 microresonator-based quantum light sources with high brightness using a subtractive wafer-scale platform.
\newblock {\em Opt. Express}, 29(16):24750--24764, Aug 2021.

\bibitem{Clemmen:09}
S.~Clemmen, K.~Phan Huy, W.~Bogaerts, R.~G. Baets, Ph. Emplit, and S.~Massar.
\newblock Continuous wave photon pair generation in silicon-on-insulator waveguides and ring resonators.
\newblock {\em Opt. Express}, 17(19):16558--16570, Sep 2009.

\bibitem{Engin:13}
Erman Engin, Damien Bonneau, Chandra~M. Natarajan, Alex~S. Clark, M.~G. Tanner, R.~H. Hadfield, Sanders~N. Dorenbos, Val Zwiller, Kazuya Ohira, Nobuo Suzuki, Haruhiko Yoshida, Norio Iizuka, Mizunori Ezaki, Jeremy~L. O'Brien, and Mark~G. Thompson.
\newblock Photon pair generation in a silicon micro-ring resonator with reverse bias enhancement.
\newblock {\em Opt. Express}, 21(23):27826--27834, Nov 2013.

\bibitem{Fujiwara:17}
Mikio Fujiwara, Ryota Wakabayashi, Masahide Sasaki, and Masahiro Takeoka.
\newblock Wavelength division multiplexed and double-port pumped time-bin entangled photon pair generation using si ring resonator.
\newblock {\em Opt. Express}, 25(4):3445--3453, Feb 2017.

\bibitem{Faruque:18}
Imad~I. Faruque, Gary~F. Sinclair, Damien Bonneau, John~G. Rarity, and Mark~G. Thompson.
\newblock On-chip quantum interference with heralded photons from two independent micro-ring resonator sources in silicon photonics.
\newblock {\em Opt. Express}, 26(16):20379--20395, Aug 2018.

\bibitem{Ma:18}
Chaoxuan Ma, Xiaoxi Wang, and Shayan Mookherjea.
\newblock Photon-pair and heralded single photon generation initiated by a fraction of a 10 gbps data stream.
\newblock {\em Opt. Express}, 26(18):22904--22915, Sep 2018.

\bibitem{Zheng_2021}
Qilin Zheng, Jiacheng Liu, Chao Wu, Shichuan Xue, Pingyu Zhu, Yang Wang, Xinyao Yu, Miaomiao Yu, Mingtang Deng, Junjie Wu, and Ping Xu.
\newblock Bright 547-dimensional hilbert-space entangled resource in 28-pair modes biphoton frequency comb from a reconfigurable silicon microring resonator.
\newblock {\em Chinese Physics B}, 31(2):024206, jan 2021.

\bibitem{Silverstone2015QubitEB}
Joshua~W. Silverstone, Raffaele Santagati, Damien Bonneau, Michael~J. Strain, Marc Sorel, Jeremy~Lloyd O'Brien, and Mark~G. Thompson.
\newblock Qubit entanglement between ring-resonator photon-pair sources on a silicon chip.
\newblock {\em Nature Communications}, 6, 2015.

\bibitem{Ma:17}
Chaoxuan Ma, Xiaoxi Wang, Vikas Anant, Andrew~D. Beyer, Matthew~D. Shaw, and Shayan Mookherjea.
\newblock Silicon photonic entangled photon-pair and heralded single photon generation with car > 12,000 and $g^{(2)}(0)$ < 0.006.
\newblock {\em Opt. Express}, 25(26):32995--33006, Dec 2017.

\bibitem{Feng:23}
Lan-Tian Feng, Yu-Jie Cheng, Xiao-Zhuo Qi, Zhi-Yuan Zhou, Ming Zhang, Dao-Xin Dai, Guang-Can Guo, and Xi-Feng Ren.
\newblock Entanglement generation using cryogenic integrated four-wave mixing.
\newblock {\em Optica}, 10(6):702--707, Jun 2023.

\end{thebibliography}






\end{document}